# COMPUTING PATHWAYS TO SYSTEMS BIOLOGY: KEY CONTRIBUTIONS OF COMPUTATIONAL METHODS IN PATHWAY IDENTIFICATION

**Sriganesh Srihari and Mark A. Ragan**

The University of Queensland, Institute for Molecular Bioscience, St. Lucia, QLD 4072 Australia.

**Abstract**

Understanding large molecular networks consisting of entities such as genes, proteins or RNAs that interact in complex ways to drive the cellular machinery has been an active focus of systems biology. Computational approaches have played a key role in systems biology by complementing theoretical and experimental approaches. Here we roadmap some key contributions of computational methods developed over the last decade in the reconstruction of biological pathways. We position these contributions in a 'systems biology perspective' to reemphasize their roles in unravelling cellular mechanisms and to understand 'systems biology diseases' including cancer.

1. Introduction

To understand the functional organisation of cells or higher biological units, often it is beneficial to conceptualize them as *systems* of interacting entities. For such a systems-level description, one needs to know (a) the entities ("parts list") that constitute the system, (b) the interactions among these entities, and (c) their dynamic behaviour under changes to internal and external conditions [1]. The goal of *systems biology* is to combine all this information into models that capture current knowledge and provide new insights and predictions about the system under previously unstudied conditions [2]. Early attempts at systems biology suffered from inadequate data to build reliable models and formulate hypotheses; however, the recent advent of high-throughput technologies has brought the initial wave of data to revive systems-level modelling and analysis, thereby seeding a revolutionary change in how biology is being studied and understood.

Understanding of complex molecular *networks* consisting of entities such as genes, proteins or RNAs connected by interactions for regulation or synthesis in cellular decision-making and responses has become a key focus of systems-level studies. Efficient computational approaches can complement theoretical and experimental approaches to model, analyse and distil knowledge from high-throughput data. The reconstruction of *biological pathways* through which cellular entities interact, signal and regulate cellular processes is certainly one of the fundamental building blocks towards understanding whole biological networks.

### *1.1 Understanding computational methods in a systems biology setting*

In order to decipher the correct and complete picture of cellular organisation, it is imperative to assess the entire collection of pathways as a whole rather than individually; the system as a whole has emergent properties that are not visible at the parts level [3]. However the reconstruction of all pathways, together with their intricate network of cross-talk and feedback loops, is a daunting task. Since pathways do not have definite start and end points or distinct boundaries, modeling them computationally is a significant challenge. Nevertheless, all computational methods developed to date model pathways as definite computable structures such as paths, trees or subnetworks. Under these circumstances, it becomes all the more crucial to place these computational methods and their contributions in a systems biology setting to assess where we stand in achieving this higher goal of systems-level understanding of cellular organisation.

In this article, we roadmap some of the key computational methods devised over the last decade for biological pathway reconstruction from high-throughput data. Although we mainly look at methods reconstructing *regulatory or signalling pathways*, this illustrates how computational methods have contributed to this area more broadly. While some methods have focused on identifying general pathways, others have specifically considered dysregulated and disease pathways, while still others have looked at additive, alternative or compensatory relationships among pathways. Further, as these methods evolved, so did their mechanisms to integrate diverse "omics" data, mainly genomics (gene expression, methylation, mutation, regulation, genetic interaction) and proteomics (protein interaction). It is both imperative and interesting to put all these developments together and consider where we stand in deciphering systems-level molecular networks and evaluate its implications.

## 2. General pathway identification

In an elegant study [4] conducted as early as 2001, Ideker et al. concentrated on a core pathway, GAL (galactose utilization), in the yeast *Saccharomyces cerevisiae*, implementing an integrated approach involving molecular expression and interactions to understand how its genes are regulated. After assembling the GAL pathway based on available information, these authors systematically perturbed each gene and measured the response through expression of a global gene set. Around the same time, large-scale protein-protein (PPI) and protein-DNA interactions were being catalogued for yeast [5], and this enabled Ideker et al. to assemble a global network of ~3000 interactions. This network was used to identify paths connecting perturbed GAL genes to every other affected gene. They identified nine genes involved in glycogen accumulation and protein metabolism, and several of unknown function, that responded strongly to galactose induction.

As more high-throughput protein interaction [5-7] and gene expression data [8] began to appear, Steffen et al. [9] in 2002 took a similar approach, assembling a large PPI network to draw possible linear paths of specified lengths starting at membrane proteins and ending on DNA-binding proteins. They scored these paths using co-expression among adjacent proteins to identify high-ranking paths corresponding to known pathways. This method, called NetSearch, enabled them to reconstruct the yeast MAPK (mitogen-activated protein kinases) pathways involved in pheromone response, filamentous growth, and maintenance of cell wall integrity. Similarly, Liu et al. [10] identified candidate sets of proteins in a PPI network, and inferred the highest-scoring order among them by measuring gene co-expression among adjacent proteins in each permutation of the set. With this scoring, they could identify the correct ordering among proteins that corresponded to MAPK pathways in yeast. In a separate attempt, Friedman et al. [11,12] used a Bayesian framework to learn expression profiles of genes perturbed in yeast mutants and used it to infer pairwise expression correlation among proteins. Through this they assembled pathways involved in purine biosynthesis and non-homologous DNA double strand break repair.

Alongside yeast, large-scale PPI data from prokaryotes including *Helicobacter pylori* [13] and eukaryotes including *Caenorhabditis elegans* [14] began to appear around 2001-2002, which enabled researchers to search in networks for pathways conserved across species. In a seminal work of this kind, Kelley et al. [15] in 2003 devised PATHBLAST, an efficient tool to align paths across multiple PPI networks. This tool enables search for homologous paths across networks by accommodating "gaps" and "mismatches". They identified 150 homologous paths of lengths ≥4 among the three species. Further, by self-matching the yeast network, Kelley et al. identified about 300 paralogous paths that they grouped into several functional pathways. Following this success, Shlomi et al. [16] proposed QPath, which improved on the results of PATHBLAST.

In the meantime, as reports [17,18] of high false-positive rates in high-throughput experiments began to surface, it became necessary to assess reliability of interactions before employing them in focused studies such as pathway identification. In a seminal study combining reliability scoring and pathway identification, Scott et al. [19] (2006) assigned a score to every interaction in the network by combining three criteria: (a) the number of times a protein pair was seen interacting in multiple experiments, (b) the Pearson correlation between expression profiles of the proteins, and (c) their small-world clustering coefficient. Using the resultant scored network, they devised two algorithms to identify pathway structures. The first identified high-scoring simple paths, while the second identified more-general structures including rooted trees and 'series-parallel' graphs. Through experiments on a yeast network of ~4500 proteins and ~14500 interactions, they successfully reconstructed several pheromone-response pathways with high accuracy.

Aided by computational tools and experimental approaches, the growth of public databases including KEGG [20] for pathways and FunCat [21] and Gene Ontology (GO) [22] for functional annotations enabled pathway identification methods to use these annotations for both prediction and validation of results. Among the first to use such diverse data was PathFinder by Bebek et al. [23] in 2007. These authors collected functional annotations from FunCat and GO for proteins in KEGG to build functional templates for pathways. They then used these templates to mine pathways from the yeast PPI network, and the high-support pathways were identified and scored using association rules mining [24]. PathFinder showed significantly better accuracy and sensitivity compared to most earlier methods, and identified several missing links among proteins in annotated pathways in databases.

In an attempt to identify general substructures beyond linear paths, Zhao et al. [25] in 2008 devised an integer linear programming (ILP)-based approach to mine the yeast PPI network. They modeled pathways as compact subnetworks between fixed starting and ending points. These subnetworks were scored using reliability scores on the edges: linear paths were scored as the sum of the edges in the paths, while general subnetworks were scored as the sum of the constituent edges. An ILP-based model was then proposed to extract high-scoring subnetworks from the PPI network. Experiments on subnetworks identified between membrane proteins and transcription factors showed that many pheromone-response pathways and signalling pathways for filamentous growth were reconstructed with high accuracy.

Liu et al. [26] (2009) noticed that most signals between proteins (for example activation, inhibition, phosphorylation, dephosphorylation and ubiquitination) were directional, so identifying the correct direction of interactions among proteins was crucial for accurate reconstruction of signalling pathways. They proposed a signal-flow model to orient interactions in the PPI network, for which they used domain interaction information among proteins. Based on the deduced orientations, they identified potential upstream-downstream relationships within protein pairs. This method successfully reconstructed several signalling pathways from the human PPI network, which matched ones annotated in KEGG and other databases.

Independent to these approaches, Boolean [27,28], Petri Nets [29] and ordinary differential equations (ODEs) [30] are a few other models that were proposed, but these have mainly focused on simulation and study of behavior of known pathways as against pathway inference from high-throughput datasets.

## 3. Exploring relationships among pathways

As methods to identify pathways improved, it became interesting to understand the relationships among different pathways and how they constituted the larger "pathway network" to regulate and govern cellular processes. This was further fuelled by the realisation that many diseases, including cancer, arise from a complex interplay between pathways acting in additive, compensatory or alternative ways to maintain aberrant behaviour of cells. While the genomics and proteomics data had already proven useful, the availability of genetic-interaction data from systematic knock-out experiments in yeast and other organisms [31-33] further aided these studies.

Among the seminal works in this direction, Kelley and Ideker [34] (2005) combined PPI and genetic-interaction (GI) networks to understand pathway relationships. They proposed that several pathways linking proteins in the PPI network were related by between-pathway interactions in the GI network. This *between-pathway model* (BPM) proposed that such pathways are involved in *compensatory* functions and buffered the loss of one another. These pathways form alternative or redundant functional groups to maintain the *robustness* of pathway mechanisms.

BPM prompted further work to look at PPI and GI interactions in an integrated manner to decipher such pathway relationships. One of the early works that took this forward was by Ulitsky and Shamir [35] (2007), who assembled a large network combining PPI and GI data, and systematically searched for BPM structures. They found that several of the pathways in KEGG [28] were related by BPM structures, indicating that these pathways functioned in a compensatory fashion. BPM relationships between proteins in different complexes with 'pivot' or shared proteins among the complexes were essential to all the host complexes. Subsequently Hescott et al. [36] (2009) integrated gene-expression data to evaluate BPM structures identified from PPI and GI networks in order to further refine redundant pathway identification.

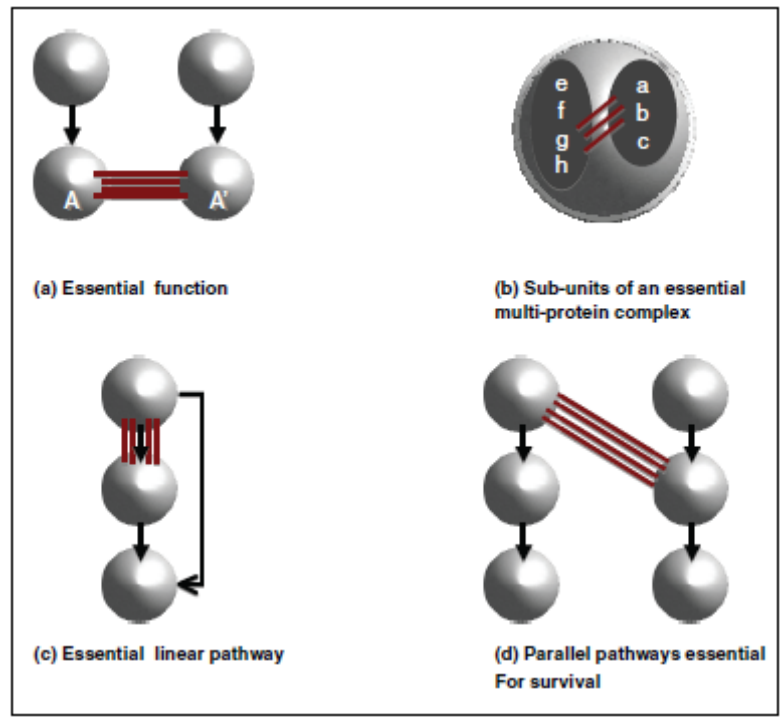

**Figure 1:** Models of synthetic lethality relationships seen between proteins within and between pathways and complexes (adapted from Le Meur and Gentleman [37] with permission from *Genome Biology*).

In this context, a subset of genetic interactions called *synthetic lethality* (SL) interactions have gained immense interest due to their prominence in connecting compensatory pathways and functions. Le Meur and Gentleman (2008) [37] analysed the enrichment of SL interactions within and between complexes and pathways; most SL interactions were between-pathway and -complexes, while a considerable number were also within these structures (see **Figure 1**). These within-complex and -pathway interactions ensured internal robustness to these structures by buffering the functions of proteins. In separate work, Ma et al. (2008) [38] proposed that finding bipartite connected subnetworks or bicliques in SL networks can help to identify groups of proteins belonging to redundant pathways. To this end, they searched for bicliques among SL interactions, and were able to identify several interesting relationships between pathways governing functions including DNA repair and DNA replication, and tubulin folding and mitosis. Brady et al. (2009) [39] further extended this approach to search for structures that they called stable bipartite subgraphs, and identified several new pairs of redundant pathway relationships.

## 4. Pathways in dysregulated functions and diseases

Many diseases, including cancer, result from dysregulated pathways and their complex interactions, and it is becoming increasingly clear that mapping these pathways is crucial to fully understand these complex diseases. This requires integrating diverse information from gene expression, mutation and regulation, protein interaction datasets and others; this has prompted integrative or "multi-omics" research towards pathway identification. Early work by Schadt et al. (2005) [40] and Tu et al. (2006) [41] focused on identifying possible linkages between (causal) mutations in DNA sequences or genes, and differentially expressed (target) genes under disease conditions. Their aim was to trace paths of differential expression from the targets back to the causal genes; these paths were hypothesised to constitute dysregulated pathways in the disease. Both groups approached this by looking for paths between target and causal genes through PPI networks. The interactions in the network were scored based on the correlation between each gene in the network and the target genes. Tu et al. simulated random walks from target genes, and the most frequently visited causal genes were evaluated for possible mutational characteristics, while the corresponding paths were analysed for involvement in disease mechanisms.

Following these approaches, several methods were devised to simulate flows to identify paths between causal and target genes. An important method by Suthram et al. [42] in 2008 used electric-circuit-based modelling [43] to simulate flows in the network. The motivation was that the random-walk-based methods were stochastic and required many simulations (about 10000 times in Tu et al.) to determine the causal gene. To propose a deterministic steady-state solution, Suthram et al. equated these random walks to the flow of electric current, and solved the network using electric-circuit theory (Kirchhoff's and Ohm's Laws). Equating the amount of current flow through each node and edge in the network to the expression level and importance of the genes, they could determine the 'true' causal genes with high accuracy.

The electric circuit method was successfully adopted by Kim et al. (2009) [44] in the study of glioblastoma multiforma. They selected a set of differentially expressed target genes that covered 158 glioblastoma cases, and then identified possible genomic loci harbouring causal genes responsible for the differential expression of target genes. By overlaying a human PPI network, Kim et al. found probable paths from target to causal genes for which they used the electric-circuit model of Suthram et al. The causal genes identified were then evaluated for the disease cases they covered. Gene Ontology-based analysis of the identified pathways showed high enrichment of processes involved in glioblastoma. Following this success, He et al. [45] used a similar approach to identify dysfunctional genes and modules in congenital heart disease (CHD).

## 5. Role of computational methods in future breakthroughs

Among the kinds of genetic interactions, synthetic lethality (SL) describes a scenario in which single-gene defects are compatible with cell viability but a combination of gene defects results in cell death [46]. In essence, these SL interactions provide *functional buffering*, sometimes described as *genetic canalisation*, that is, buffering of pathways against the tendency of new alleles or mutated genes to make non-optimal phenotypes [47]. With the assembling of major pathways involved in core cellular processes affected in cancer, including DNA replication, DNA damage repair and cell-cycle checkpoints, it is now clear that buffering among pathways maintains viability in cancerous cells, potentially weakening anti-cancer therapies aimed at blocking individual pathways. However on the positive side, this hints that "sweet spots" capable of overcoming such compensatory arrangements can be identified that may lead to effective anti-cancer strategies in the future [48].

The PARP/BRCA relationship has become the poster child for SL-based cancer research today [48,49]. BRCA1 is a large protein expressed during the S and G2 phases of the cell cycle, and involved in the DNA double-strand break (DSB) repair pathway, one of several pathways required for maintaining DNA integrity. BRCA1-mutant cells have a defect in DNA damage repair, specifically in homologous recombination (HR)-mediated repair. In normal cells, the loss of BRCA1 activity is sensed in S-phase, resulting in immediate TP53-mediated cell death. However, tumor cells acquire a state in which repeated transit through S-phase can be accomplished despite loss of BRCA1 function. This leads to genomic instability and potential mutations in other crucial genes including p53 required to control cell proliferation. A major breakthrough in BRCA1-mutant cancers was heralded by the finding that BRCA1 mutant cells are sensitive to PARP inhibitors. PARP1 is involved mainly in the DNA single-strand break (SSB) repair pathway. In the context of PARP inhibition, unrepaired SSBs accumulate into DSB equivalents upon entry into S-phase. In normal cells, these lesions are repaired by the HR-mediated DSB repair pathways. However, in the absence of BRCA1 there may be greater reliance on PARP-mediated pathways, failing which DNA DSBs accumulate and lead to cell arrest and death. In essence, inhibition of PARP deals a double blow to BRCA1-deficient cells leading to cell death. Several PARP-inhibitor compounds are now under clinical or pre-clinical trials for use in anti-cancer therapy based on this concept [48].

Although the PARP/BRCA relationship is only a conditional one and not a panacea, it reflects the extent and kind of intricate relationships that need to be deciphered to map and understand the "pathway network", and therefore to understand diseases like cancer. As Laubenbacher et al. [3] rightly put it as, "cancer is a systems biology disease". While our current knowledge of pathways in cancer is still incomplete, immense efforts are underway to identify new players (genes, proteins and whole pathways) as well as to implicate existing ones in new roles – for example, the recent (2009) implication of a SUMO-mediated pathway in the BRCA1 response to genotoxic stress [50].

Computational methods have a key role to play alongside experimental approaches. As an example, in recent remarkable research Rodriguez et al. [51] (2012) constructed a Boolean network model of the Fanconi anaemia/breast cancer (FA/BRCA) pathway to simulate the interstrand cross-links (ICL) repair process, whose inhibition is known to result in a chromosomal instability syndrome called Fanconi anaemia. Rodriguez et al. modelled knowledge of ICL as logical rules, obtaining a Boolean network of 28 nodes and 122 regulatory interactions. These Boolean rules captured relationships among genes and pathways; for example, ICL can be responded to either by generating a DSB that is subsequently repaired by the BRCA1-mediated HR pathway, OR by bypassing this with the help of proliferation cell nuclear antigen (PCNA) and translesion synthesis (TLS) and repairing it with the nuclear excision (NER) pathway. Next, by fixing the loss- and gain-of-function mutants as 0 or 1 in the network, they performed dynamical simulations to understand the state of the network, that is, the alternative pathways favoured under various mutational conditions. In this way they inferred key buffering mechanisms that may compensate for the defective FA/BRCA pathway, which are worth further research and experimental validation.

## 6. Conclusions

In order to decipher a correct and complete picture of cellular organisation, it is necessary to take a systems-level view. Mapping the molecular network is a crucial step toward this goal. However, mapping the entire network at one go can be daunting (and the current data are inadequate to support this), so mapping individual pathways is a rational way to proceed. Having said that, knowledge of individual pathways must be reflected back to the systems-level context, lest we miss the bigger picture.

Computational approaches have played a key role in identification and mapping of pathways. In this article, by drawing a roadmap of key contributions from computational approaches, and describing instances in which collective understanding of multiple pathways is necessary (for example, in cancer), we have attempted to put all these developments in a *systems-level perspective*. Future breakthroughs in understanding cellular mechanisms and diseases can come only by taking such a systems-level view. To this end, computational approaches will continue to play a key role.

**Key Points**

- Mapping the network of genes, proteins, RNA, and other molecules is a crucial step towards realizing the 'systems biology goal' of understanding cellular organisation.

- Computational methods play a key role in this by complementing theoretical and experimental approaches.

- We have highlighted key contributions of computational approaches in identification of signalling/regulatory pathways. These approaches have gone hand-in-hand with the improvements in high-throughput techniques, and have integrated diverse "omics" datasets.

- Through these contributions, we reemphasize that computational methods have a key role in future breakthroughs in understanding cellular organization and also complex systems-level diseases like cancer.

# References

[1] Kitano, H. Systems Biology: A brief overview. Science 2002; 295:1662--1664.

[2] Albert, R. Network inference, analysis, and modeling in systems biology. Plant Cell 2007; 19:3327—3338.

[3] Laubenbacher, R., Hower, V., Jarrah, A., Torti, S.V., Shulaev, V., Mendes, P., Torti, M., Akman, S. A systems biology view of cancer. Biochim Biophys Acta 2009; 1796(2):129—139.

[4] Ideker, T., Thorsson, V., Ranish, J., Christmas, R., Buhler, J., Eng, J.K., Bumgarner, R., Goodlett, D., Aebersold, R., Hood, L. Integrated genomic and proteomic analyses of a dystematically perturbed metabolic network. Science 2001; 292(5518):929--934.

[5] Schwikowski, B., Uetz, P., Fields, S. A network of protein-protein interactions in yeast. Nature Biotechnol. 2000; 18:1257--1261.

[6] Uetz, P., Giot, L., Cagney, G., Mansfield, T., Judson, R., Knight, J., Lokshon, D., Narayan, V., Srinivasan, M., Pochart, P. A comprehensive analysis of protein-protein interactions in *Saccharomyces cerevisiae.* Nature 2000; 403:623--627.

[7] Ito, T., Chiba, T., Ozawa, R., Yoshida, M., Hattori, M., Sakaki, Y. A comprehensive two-hybrid analysis to explore the yeast protein interactome. Proc Natl Acad Sci 2001; 98:4569--4574.

[8] Hughes, T.R., Marton, M.J., Jones, A., Roberts, C.J., Stoughton, R., Armour, C., Bennett, H., Coffey, E., Dai, H., He, Y. Functional discovery via a compendium of expression profiles. Cell 2000; 102:109--126.

[9] Steffen, M., Petti, A., Aach, J., D'haeseleer, P., Church, G. Automated modelling of signal transduction networks. BMC Bioinformatics 2002; 3:34.

[10] Liu, Y., Zhao, H. A computational approach for ordering signal transduction pathway components from genomics and proteomics data. BMC Bioinformatics 2004; 5:158.

[11] Friedman, N., Linial, M., Nachman, I., Pe'er, D. Using Bayesian networks to analyse expression data. J Computational Biology 2000;7(3-4):601--620.

[12] Pe'er, D., Regev, A., Elidan, G., Friedman, N. Inferring subnetworks from perturbed expression profiles. Bioinformatics 2001; 17(Suppl 1):215--224.

[13] Rain, J., Selig, L., de Reuse, H., Battaglia, V., Reverdy, C., Simon, C., Lenzen, G., Petel, F., Wojcik, J., Schachter, V., Chemama, Y., Labigne, A., Legrain, P. The protein-protein interaction map of *Helicobacter pylori*. Nature 2001; 409(6817):211--215.

[14] Boulton, S.J., Gartner, A., Reboul, J., Vaglio, P., Dyson, N., Hill, DE., Vidal, M. Combined functional genomic maps of the C. *elegans* DNA damage response. Science 2002; 295(5552):127--131.

[15] Kelley, B.P., Sharan, R., Karp, R.M., Sittler, T., Root, D., Stockwell, B., Ideker, T. Conserved pathways within bacteria and yeast as revealed by global protein network alignment. Proc. Natl. Acad. Sci. 2003; 100(20):11394--11399.

[16] Shlomi, T., Segal, D., Ruppin, E., Sharan, R. QPath: a method for querying pathways in a protein-protein interaction network. BMC Bioinformatics 2006; 7:199.

[17] von Mering, C., Krause, R., Snel, B., Cornell, M., Oliver, S.G., Fields, S., Bork, P. Comparative assessment of large-scale data sets of protein-protein interactions, Nature 2002; 417:399--403.

[18] Bader, G., Chaudhuri, A., Rothberg, J., Chant, J. Comparative assessment of large-scale data sets of protein–protein interactions. Nature Biotechnol. 2004; 22(1):78--85.

[19] Scott, J., Ideker, T., Karp, R.M., Sharan, R. Efficient algorithms for detecting signaling pathways in protein interaction networks. J Comp Biology 2006; 13(2):133--144.

[20] Kanehisa M., Goto S. KEGG: Kyoto Encyclopedia of Genes and Genomes. Nucleic Acids Res 2000; 28:27-30.

[21] Ruepp A, Zollner A, Maier D, Albermann K, Hani J, Mokrejs M, Tetko I, Guldener U, Mannhaupt G, Munsterkotter M, Mewes HW. The FunCat, a functional annotation scheme for systematic classification of proteins from whole genomes. Nucleic Acids Res 2004; 32(18):5539-5545.

[22] Ashburner M, Ball CA, Blake JA, Botstein D, Butler H, Cherry JM, Davis AP, Dolinski K, Dwight SS, Eppig JT, Harris MA, Hill DP, Issel-Tarver L, Kasarskis A, Lewis S, Matese JC, Richardson JE, Ringwald M, Rubin GM, Sherlock G. Gene ontology: tool for the unification of biology. Nat Genetics 2000; 25:25--29.

[23] Bebek, G., Yang, J. PathFinder: mining signal transduction pathway segments from protein-protein interaction networks. BMC Bioinformatics 2007; 8:335.

[24] Agarwal, R., Srikanth, R. Fast algorithms for mining association rules. Proceedings of 20th International Conference on Very Large Databases (VLDB) 1994; 487--499.

[25] Zhao, X-M., Wang, R-S., Chen, L., Aihara, K. Uncovering signal transduction networks from high-throughput data by integer linear programming. Nucleic Acids Res 2008; 36(9):e48.

[26] Liu, W., Li, D., Wang, J., Xie, H., Zhu, Y., He, F. Proteome-wide prediction of signal flow direction in protein interaction networks based on interacting domains. Mol Cell Prot 2009; 8(9):2063--3070.

[27] Kauffman, S.A. Metabolic stability and epigenesis in randomly constructed gene sets. J. Theoretical Biology 1969; 22:437--467.

[28] Thomas, R., D'Ari, R. Biological feedback. Boca Raton, FL CRC Press, 1990.

[29] Peterson, J.L. Petri Nets. ACM J. Computing Surveys 1977; 9(3):223--252.

[30] Hughey, J.J., Lee, T.K., Covert, M.W. Computational modeling of signalling networks. Wiley Interdiscip Rev Syst Biol Med 2010; 2(2):194-209.

[31] Tong, AHY., Lesage, G., Bader, GD., Ding, H., Xu, H., et al. Global mapping of the yeast genetic interaction network. Science 2004; 303:808--813.

[32] Boone, C., Bussey, H., Andrews, BJ. Exploring genetic interactions and networks with yeast. Nat Rev Genet 2007; 8:437--449.

[33] Koh, J., Ding, H., Costanzo, M., Baryshnikova, A., Toufighi, K., Bader, G., Myers, C., Andrews, BJ., Boone, C. DRYGIN: a database of quantitative genetic interaction networks in yeast. Nucleic Acids Res 2010; 38:D502-D507.

[34] Kelley, R., Ideker, T. Systematic interpretation of genetic interactions using protein networks. Nat Biotechnol 2005; 23(5):561.

[35] Ulitsky, I., Shamir, R. Pathway redundancy and protein essentiality revealed in the Saccharomyces cerevisiae interaction networks. Mol Sys Biol, 2007; 3:104.

[36] Hescott, B., Leiserson, M., Cowen, L., Slonim, D. Evaluating between-pathway models with expression data. Proceedings of the 13th International Conference on Research in Computational Molecular Biology (RECOMB) 2009; pp.372--385.

[37] Le Meur, N., Gentleman, R. Modeling synthetic lethality. Genome Biology 2008; 9:R135.

[38] Ma, X., Tarone, A.M., Li, W. Mapping genetically compensatory pathways from synthetic lethal interactions in yeast. PLoS ONE 2008; 3(4):e1992.

[39] Brady, A., Maxwell, K., Daniels, N., Cowen, L. Fault tolerance in protein interaction networks: stable bipartite subgraphs and redundant pathways. PLoS ONE 2009; 4(4) :e5364.

[40] Schadt, E., Lamb, J., Yang, X., Zhu, J., Edwards, S., Guhathakurta, D., Sieberts, S.K., Monks, S., Reitman, M., Zhang, C., et al. An integrative genomics approach to inder causal associations between gene expression and disease. Nature Genetics 2005; 37:710—717.

[41] Tu, Z., Wang, L., Artbeitman, M.N., Chen, T., Sun, F. An integrative approach for causal gene identification and gene regulatory pathway inference. Bioinformatics 2006; 22(14):e489--e496.

[42] Suthram, S., Beyer, A. Karp, R., Eldar, Y., Ideker, T. eQED: an efficient method for interpreting eQTL associations using protein networks. Mol Sys Biol 2008; 4:162.

[43] Odum, E. Fundamentals of ecology. W.B.Saunders, Philadelphia 1953.

[44] Kim, Y-A., Wutchy, S., Przytycka, TM. Identifying causal genes and dyregulated pathways in complex diseases. PLoS Comp Bio 2009; 7(3):e1001095.

[45] He, D., Liu, Z-P., Chen, L. Identification of dysfunctional modules and disease genes in congenital heart disease by a network-based approach. BMC Genomics 2011; 12:592.

[46] Ashworth, A., Lord, C.J., Ries-Filho, J. Genetic interactions in cancer progression and treatment. Cell 2011; 145.

[47] Hartman, J., Garvik, B., Hartwell, L. Principles for the buffering of genetic variation. Science 2001; 291:1001--1004.

[48] Lord, C.J., Ashworth, A. The DNA damage response and cancer therapy, Nature 2012; 481:287--294.

[49] Amal, A., Ganesan, S. BRCA1, PARP, and 53BP1: conditional synthetic lethality and synthetic viability. J Mol Cell Biol 2011; 3:66--74.

[50] Morris, JR., Boutell, C., Keppler, M., Densham, R., Weekes, D., Alamshah, A., Butler, L., Galanty, Y., Pangon, L., Kuichi, T., Ng, T., Solomon, E. The SUMO modification pathway is involved in the BRCA1 response to genotoxic stress. Nature 2009; 462(7275):886—890.

[51] Rodriguez, A., Sosa, D., Torres, L., Molina, B., Fras, S., Mendoza, L. A Boolean network model of the FA/BRCA pathway. Bioinformatics 2012; 28(6):858--866.